# A Design Technique for Faster Dadda Multiplier

B. Ramkumar, V. Sreedeep and Harish M Kittur, *Member, IEEE*

*Abstract*- In this work faster column compression multiplication has been achieved by using a combination of two design techniques: partition of the partial products into two parts for independent parallel column compression and acceleration of the final addition using a hybrid adder proposed in this work. Based on the proposed techniques 8, 16, 32 and 64-bit Dadda multipliers are developed and compared with the regular Dadda multiplier. The performance of the proposed multiplier is analyzed by evaluating the delay, area and power, with 180 nm process technologies on interconnect and layout using industry standard design and layout tools. The result analysis shows that the 64-bit regular Dadda multiplier is as much as 41.1% slower than the proposed multiplier and requires only 1.4% and 3.7% less area and power respectively. Also the power-delay product of the proposed design is significantly lower than that of the regular Dadda multiplier.

*Index Terms*- Column compression, Dadda multiplier, Faster, Hybrid final adder.

## I. INTRODUCTION

High speed multiplication is a primary requirement of high performance digital systems. In recent trends the column compression multipliers are popular for high speed computations due to their higher speeds [1-2]. The first column compression multiplier was introduced by Wallace in 1964 [3]. He reduced the partial product of *N* rows by grouping into sets of three row set and two row set using (3,2) counters and (2,2) counters respectively. In 1965, Dadda altered the approach of Wallace by starting with the exact placement of the (3,2) counters and (2,2) counters in the maximum critical path delay of the multiplier [4]. Since 2000's, a closer reconsideration of Wallace and Dadda multipliers has been done and proved that the Dadda multiplier is slightly faster than the Wallace multiplier and the hardware required for Dadda multiplier is lesser than the Wallace multiplier [5-6]. Since the Dadda multiplier has a faster performance, we implement the proposed techniques in the same and the improved performance is compared with the regular Dadda multiplier.

The column compression multipliers have total delays that are proportional to the logarithm of the operand word lengths which is unlike the array multipliers which have speeds proportional to the word length [7-8]. The total delay of the multiplier can be split up into three parts: due to the Partial Product Generation (PPG), the Partial Product Summation Tree (PPST), and finally due to the Final Adder [9]. Of these the dominant components of the multiplier delay are due to the PPST and the final adder. The relative delay due to the PPG is small. Therefore significant improvement in the speed of the multiplier can be achieved by reducing the delay in the PPST and the final adder stage of the multiplier. In this work the delay introduced by the PPST is reduced by using two independent structures in the partial products. The proposed hybrid final adder computes the final products much faster.

This paper is structured as follows: Sections II and III describe the design of parallel structures for the PPST and the design of hybrid final adder structure respectively. Section IV reports the ASIC implementation details and the simulation results. Finally, Section V summarizes the analysis. Throughout the paper, it is assumed that the number of bits in the multiplier and multiplicand are equal.

## II. DESIGN OF PARALLEL STRUCTURES

The multiplication process begins with the generation of all partial products in parallel using an array of AND gates. The next major steps in the design process are partitioning of the partial products and their reduction process. Each of these steps are elaborated in the following subsections.

### A. Partitioning the partial products

We consider two *n*-bit operands $a_{n-1}a_{n-2}...a_2a_1a_0$ and $b_{n-1}b_{n-2}...b_2b_1b_0$ for *n* by *n* Baugh-Wooley multiplier, the partial products of two *n*-bit numbers are $a_ib_j$ where *i,j* go from 0,1,..*n*-1. The partial products form a matrix of *n* rows and 2*n*-1 columns as show in Fig. 1(a). To each partial product we assign a number as shown in Fig. 1 (a), e.g. $a_0b_0$ is given an index 0, $a_1b_0$ the index 1 and so on. For convenience we rearrange the partial products as shown in Fig 1(b). The longest column in the middle of the partial products contributes to the maximum delay in the PPST.

Therefore in this work we split-up the PPST into two parts as shown in the Fig. 1(c), in which the Part0 and part1 consists of *n* columns. We then proceed to sum up each column of the two parts in parallel. The summation procedure adopted in this work is described in the next section.

### B. The Dadda based reduction

Next the partial products of each part are reduced to two rows by the using (3,2) and (2,2) counters based on the regular Dadda reduction algorithm as shown in Fig. 2 and Fig. 3. The grouping of 3-bits and 2-bits indicates (3,2) and (2,2) counters respectively and the different colors classify the difference between each column, where *s* and *c* denote *partial sum* and *partial carry* respectively. E.g. the bit positions of *6* and *13* in part0 are added using a (2,2) counter to generate sum *s0* and *c0*. The *c0* is carried to the next column where it is to be added up with the sum *s1* of a (3,2) counter adding *7, 14* and *21*. The carry *c1* of (3,2) counter is added to the next column. The final two rows of each part are summed using a Carry Look-ahead Adder (CLA) to form the partial final products of a height of one bit column which indicated at the bottom of Fig. 2 and Fig. 3.

Fig. 1. Partitioning the partial products: (a) Partial product array diagram for 8*8 multiplier, (b) An Alternative Representation, (c) Partitioned structure of multiplier showing part0 and part1.

Fig. 2. Reduction of the partial products of part1 based on the Dadda approach.

The two parallel structures for Fig. 2 and Fig. 3 based on the Dadda approach are shown in Fig. 4, where HA, FA, p0, p1 and p denote Half Adder ((2,2)counter)), Full Adder ((3,2)counter), partial final product from part0, partial final product from part1 and final product respectively. The numerals residing on the HA and FA indicates the position of partial products. The output of part0 and part1 are computed independently in parallel and those values are added using a high speed hybrid final adder to get the final product.

However, before we proceed to carry out the final addition with the proposed hybrid adder, we first carry out the final addition with the CLA for both the unpartitioned Dadda multiplier and the partitioned Dadda multiplier. This enables us to evaluate and analyze the effect of partitioning the PPST into two parts. The simulation results are listed in Table I and Table II. The comparison between the Table I and Table II gives that the percentage improvement in delay, area and power of the partitioned multipliers with respect to the regular Dadda multiplier.

It can be seen that for the 8-bit multiplier, there is no improvement in the speed, area and power. But with the increase in the word size, the improvement in the speed, area and power of the partitioned multipliers increases. There is a maximum of 10.5% improvement in delay for the 64-bit multiplier with only a slight increase in the area and power of 1% and 1.8% respectively.

Having clearly demonstrated the reduction in the delay of the Dadda multipliers due to the partitioning of the partial products we now proceed to further enhance the speed of the proposed multiplier. The further improvement in the performance can be achieved by replacing the CLA with the proposed hybrid final adder structure which is elaborated in the next section.

| | | | | | | |
|---|---|---|---|---|---|---|
| 63 | 55 | 47 | 39 | 31 | 23 | 15 |
| | 62 | 54 | 46 | 38 | 30 | 22 |
| | | 61 | 53 | 45 | 37 | 29 |
| | | | 60 | 52 | 44 | 36 |
| | | | | 59 | 51 | 43 |
| | | | | | 58 | 50 |
| | | | | | | 57 |
| 63 | 55 | 47 | 39 | 31 | s24 | s23 |
| | 62 | 54 | 46 | c24 | c23 | 29 |
| | | 61 | 53 | 38 | 37 | 36 |
| | | | 60 | 45 | 44 | 43 |
| | | | | 52 | 51 | 50 |
| | | | | 59 | 58 | 57 |
| 63 | 55 | 47 | s28 | s27 | s26 | s25 |
| | 62 | c28 | c27 | c26 | c25 | 43 |
| | | 54 | 60 | s30 | s29 | 50 |
| | | 61 | c30 | c29 | 58 | 57 |
| 63 | 55 | s33 | s32 | s31 | s30 | s29 |
| | c33 | c32 | c31 | c30 | c29 | 50 |
| | 62 | 61 | c30 | c29 | 58 | 57 |
| 63 | s39 | s38 | s37 | s36 | s35 | s34 |
| c39 | c38 | c37 | c36 | c35 | c34 | 57 |

p1[15] p1[14] p1[13] p1[12] p1[11] p1[10] p1[9] p1[8]

Fig. 3. Reduction of multiplier partial products of part2 based on the Dadda reduction tree.

## III. THE HYBRID FINAL ADDER DESIGN

In previous works the hybrid final adder designs used to achieve the faster performance in parallel multipliers were made up of CLA (Carry Lookahead Adder) and CSLA (Carry Select Adder) [9-11]. But due to the structure of the CSLA, it occupies more chip area than other adders. Thus to achieve the optimal performance, the proposed hybrid adder in this work uses MBEC (Multiplexers with Binary to Excess-1 Converters) and Ripple Carry Adder (CLA) for fast summation of uneven input arrival time of the signals originating from the PPST. The MBEC adder provides faster performance than Carry Save Adder (CSA) and Carry Look Ahead (CLA) adder [12]. Also it consumes less area and power than the Carry Select Adder (CSLA) [13].

### A. Hybrid Adder for 8 by 8 Multiplier

Once each part of the partial products has been reduced to a height of one bit column, we get the final partial products as follows,

p0[10] p0[9] p0[8] p[7] p[6] p[5] p[4] p[3] p[2] p[1] p[0]
p1[15] p1[14] p1[13] p1[12] p1[11] p1[10] p1[9] p1[8]

The p0[10:8] are the exceeding carry bits of part0 and p1[15] is the carry bit of part1. The p[7:0] of part0 are directly assigned as the final products. To find the remaining p[15:8], we use the CLA and the MBEC shown in Fig. 5.

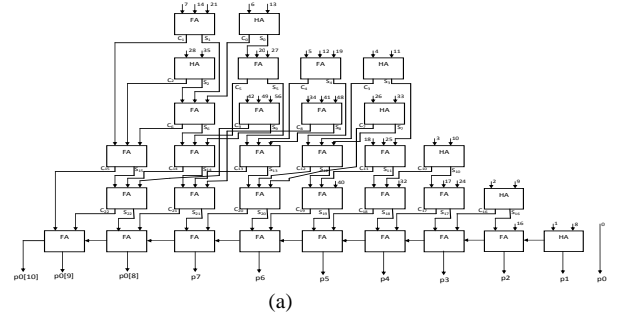

(a)

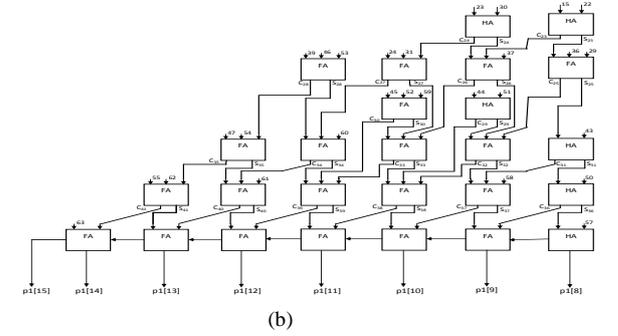

(b)

Fig. 4. The Dadda based implementation: (a) Implementation of part1, (b) Implementation of part2

TABLE I
REGULAR DADDA MULTIPLIER WITH CLA

| Multiplier N by N | Area ($\mu m^2$) | Delay (ns) | Power ($\mu W$) |
|---|---|---|---|
| 8 by 8 | 8,428 | 3.40 | 6.32 |
| 16 by 16 | 29,169 | 4.71 | 33.09 |
| 32 by 32 | 105,237 | 5.92 | 210.50 |
| 64 by 64 | 397,146 | 7.54 | 925.92 |

TABLE II
PARTITIONED DADDA MULTIPLIER WITH CLA

| Multiplier N by N | Area ($\mu m^2$) | Delay (ns) | Power ($\mu W$) |
|---|---|---|---|
| 8 by 8 | 8,957 | 3.51 | 6.85 |
| 16 by 16 | 30,241 | 4.61 | 35.22 |
| 32 by 32 | 107,362 | 5.47 | 218.76 |
| 64 by 64 | 386,629 | 6.94 | 952.59 |

The p0[10:8] and p1[10:8] are added using 3-bit CLA which finds p[10:8]. To obtain the remaining p[15:11], the p1[15:11] are assigned to the input of 5-bit MBEC, which produce the two partial results p1[15:11] with *Cin* of '0' and the 5-bit BEC output with the *Cin* of '1'. Depending on the *Cout* of CLA(c[10]), the mux provides the final p[15:11] without having to ripple the carry through p1[15:11].

The 8-bit multiplier uses a single 5-bit MBEC in the final adder. But the large bit sized multipliers requires multiple MBEC and each of them requires the selection input from the carry output of the preceding MBEC. Therefore to generate the carry output from the MBEC, an additional block is developed which is called MBECWC (MBEC With Carry). The detailed structures of the 5-bit BEC without carry (BEC) and with carry (BECWC) are shown Fig. 6(a) and Fig. 6(b). The BEC gets *n* inputs and generates *n* output; the BECWC gets *n* input and generates *n+1* output to give the carry output as the selection input of the next stage *mux* used in the final adder design of 16, 32 and 64-bit multipliers. The function table of BEC and BECWC are shown in Table III.

## B. Variable Block Hybrid Adder

The variable size of adder blocks always leads to faster adders than fixed size block adder [14]. Thus to further improve the speed of addition, we breakdown the ripple of gates in the MBEC into multiple size groups of size $2^n$, where n $\geq$ 2. Based on this approach the final adder design for 16, 32 and 64-bit multipliers are shown in Fig. 7. In MBECWC, the mux is getting *n*-bits of data input "as it is" input for selection input '0' and *n+1*-bits of data input from the BECWC output for selection input '1'. Thus to make equal the size of the inputs to the mux, the one bit '0' is appended as the MSB (Most Significant Bit) to the *n*-bits of input. E.g. In Fig. 7(a), the 10:5 mux of MBECWC gets the two inputs: 4-bits (*n*-bits) of p[23:20] for selection input '0' and 5-bits (*n+1*-bits) from the 4-bit BECWC for selection input '1' respectively. Thus to make equal the size of the inputs, the one bit '0' is appended as the MSB to the input of p[23:20] is like {0,p[23:20]}.

TABLE III
FUNCTION TABLE OF 5-BIT BEC & BECWC

| Input | BEC without carry | BEC with carry | |
|---|---|---|---|
| b[4:0] | x[4:0] | cy | x[4:0] |
| 00000 | 00001 | 0 | 00001 |
| 00001 | 00010 | 0 | 00010 |
| 00010 | 00011 | 0 | 00011 |
| 00011 | 00100 | 0 | 00100 |
| 00100 | 00101 | 0 | 00101 |
| ⋮ | ⋮ | ⋮ | ⋮ |
| 11011 | 11100 | 0 | 11100 |
| 11100 | 11101 | 0 | 11101 |
| 11101 | 11110 | 0 | 11110 |
| 11110 | 11111 | 0 | 11111 |
| 11111 | 00000 | 1 | 00000 |

To analyze independently the effect of the proposed hybrid adder, the partitioned multiplier with CLA final adder is compared with the partitioned multiplier along with the proposed hybrid adder. The simulation results are listed in Table IV and Table V. The comparison between the Table IV and Table V gives that the percentage improvement in the delay, area and power of the proposed multiplier (partitioned multiplier with hybrid final adder) with respect to the partitioned multiplier with CLA final adder.

The plot clearly shows that the performance improvement in delay increases with the word size of the multiplier. The speed of the 8, 16, 32 and 64-bit multipliers are improved 14.9%, 21.1%, 25.2% and 27.7% respectively. The area and power overhead for all word sizes is only slightly higher.

## IV. ASIC IMPLEMENTATION AND SIMULATION RESULTS

The ASIC implementation of proposed design follows the cadence design flow. The design has been developed using Verilog-HDL and synthesized in Encounter RTL compiler using typical libraries of TSMC 180nm technology. The Cadence SoC Encounter is adopted for Placement & Routing (P&R) [15]. Parasitic extraction is performed using

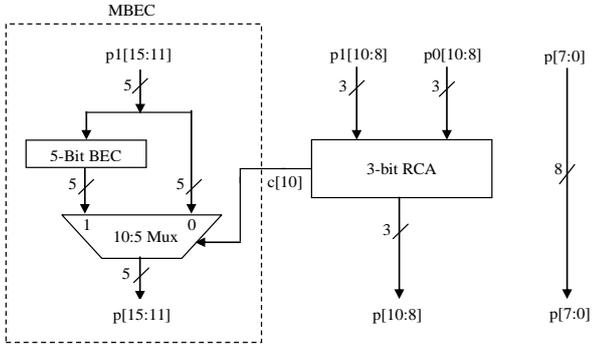

Fig. 5. Hybrid final adder of 8 by 8 multiplier

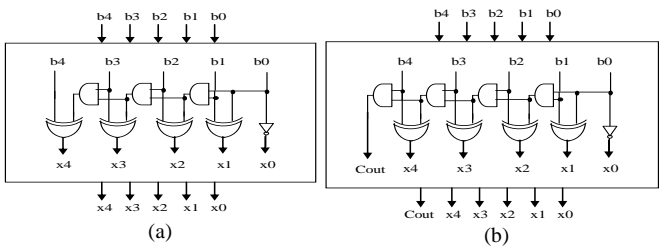

Fig. 6. The 5-bit Binary to Execss-1 Code Converter: (a) BEC (without carry), (b) BECWC (with carry).

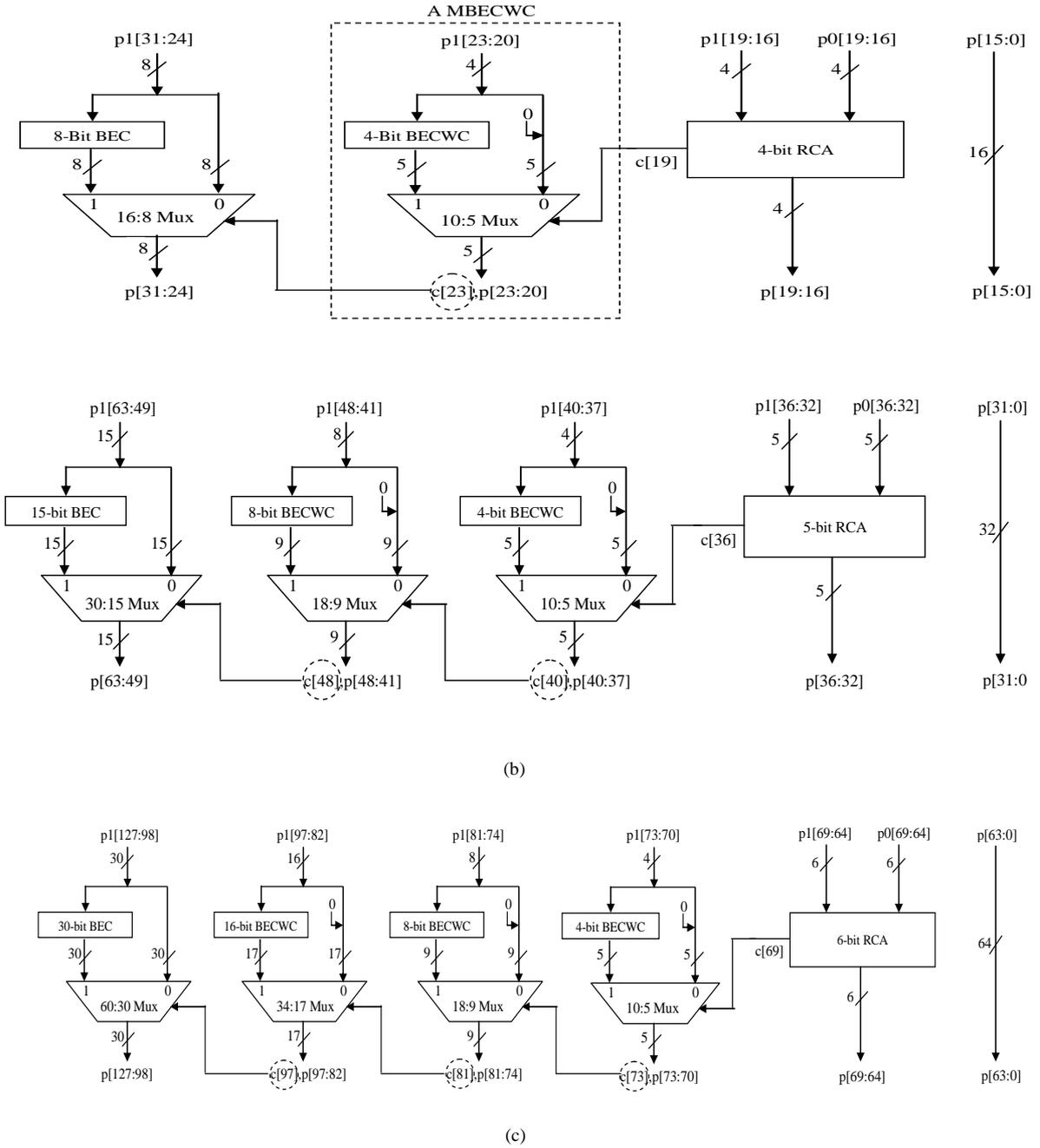

Fig. 7. Variable block hybrid final adder: (a) For 16-bit multiplier, (b) For 32-multiplier, (c) For 64-bit multiplier.

TABLE IV
PARTITIONED DADDA MULTIPLIER WITH CLA

| Multiplier $N$ by $N$ | Area ($\mu m^2$) | Delay ($ns$) | Power ($\mu W$) |
|---|---|---|---|
| 8 by 8 | 8,957 | 3.51 | 6.85 |
| 16 by 16 | 30,241 | 4.61 | 35.22 |
| 32 by 32 | 107,362 | 5.47 | 218.76 |
| 64 by 64 | 386,629 | 6.94 | 952.59 |

TABLE V
PARTITIONED DADDA MULTIPLIER WITH HYBRID ADDER

| Multiplier $N$ by $N$ | Area ($\mu m^2$) | Delay ($ns$) | Power ($\mu W$) |
|---|---|---|---|
| 8 by 8 | 9,144 | 3.38 | 7.07 |
| 16 by 16 | 30,577 | 4.13 | 35.99 |
| 32 by 32 | 107,491 | 4.71 | 221.01 |
| 64 by 64 | 381,776 | 5.51 | 966.45 |

Encounter Native RC extraction tool. The extracted parasitic RC (SPEF format) is back annotated to Common Timing Engine in Encounter Platform for static timing analysis. For each word size of the multiplier, the same VCD (Value Changed Dump) file is generated for possible input conditions and imported the same to Cadence Encounter. Power Analysis to perform the power simulations. The similar design flow is followed for both the designs in this work.

## V. RESULT SUMMARY

The comparison between the Table I (regular Dadda multiplier with CLA) and Table V (partitioned multiplier with hybrid adder) summarizes the enhanced performance of the proposed multiplier in terms of percentages which are listed in Table VI. It exhibits that the area of the regular Dadda multiplier is only slightly lesser, ranging from 7.7% to 1.4% for the 8, 16, 32 and 64-bits respectively, than the area of the proposed multiplier. It is clear that the area overhead of the proposed multiplier continuously decreases with increasing word size and is only 1.4% for the 64-bit multiplier.

The power consumption of the regular Dadda multiplier is 5.2% less than the proposed multiplier for the 8-bit word size. With increasing word size the difference in power requirement of the proposed and the Dadda multiplier decreases. Thus the 64-bit Dadda multiplier requires only 3.7% less power than the proposed multiplier.

The delay values clearly indicate that the proposed multiplier is always faster than the regular Dadda multiplier, also with increasing word size the percentage reduction of the delay increases. The speed enhancement is significant for the 64-bit where the regular Dadda requires 41.1% more time than the proposed multiplier.

## VI. CONCLUSION

We have successfully achieved faster multiplication by using a combination of two design techniques; partitioning of the partial products into two parts to perform independent parallel column compression and fast final addition using hybrid final adder structure. The result analysis shows that the power and area overheads are not significant. But the speed and power-delay product improvements are significant compared to the regular Dadda multipliers. The proposed multiplier design technique can be implemented with any type of parallel multipliers to achieve faster performance.

**B.Ramkumar** received the B.E. degree in Electronics and Communication Engineering from the Madurai Kamaraj University, Madurai, in the year 2004, and the M.E. degree in VLSI design from the Anna University, Chennai, in 2006. Currently, he is pursuing Ph.D at the VIT University, Vellore.

**V.Sreedeep** received the B.Tech. degree in Electrical and Electronics Engineering from the Jawaharlal Nehru Technological University, Anantapur, in the year 2004. Currently, he is pursuing M.Tech at the VIT University, Vellore.

**Harish M Kittur** received the B. Sc. degree in Physics, Mathematics and Electronics from the Karnataka University, Dharwad, in 1994. M. Sc. in Physics from the Indian Institute of Technology, Mumbai, in 1996. M.Tech. in Solid State Technology in the year 1999 from Indian Institute of Technology, Madras, and Ph. D. in Physics from the RWTH Aachen in the year 2004. He is a member of IEEE and IETE.

TABLE VI
PERFORMANCE OF THE REGULAR WITH REFERENCE TO THE PROPOSED DADDA MULTIPLIER

| Multiplier $N$ by $N$ | Area % | Delay % | Power % |
|---|---|---|---|
| 8 by 8 | -8.5 | + 0.5 | -11.8 |
| 16 by 16 | -4.8 | + 12.21 | -8.76 |
| 32 by 32 | -2.1 | + 20.40 | -4.99 |
| 64 by 64 | 3.8 | + 26.91 | -2.21 |